\documentclass[twocolumn,showpacs,preprintnumbers,amsmath,amssymb]{revtex4}

\usepackage{graphicx}
\usepackage{dcolumn}
\usepackage{bm}

\begin{document}

\preprint{}

\title{Site-Dilution in quasi one-dimensional antiferromagnet 
Sr$_2$(Cu$_{1-x}$Pd$_x$)O$_3$: reduction of N\'eel Temperature
and spatial distribution of ordered moment sizes}

\author{K. M. Kojima}\email{kojima@phys.s.u-tokyo.ac.jp}
\author{J.~Yamanobe}
\author{H.~Eisaki}
\author{S. Uchida}%
\affiliation{%
Department of Physics, University of Tokyo, 
Hongo 7-3-1, Bunkyo, Tokyo 113-0033, Japan
}%
\author{Y.~Fudamoto}
\author{I.M.~Gat}
\author{M.I.~Larkin}
\author{A.~Savici}
\author{Y.J.~Uemura}
\affiliation{%
Department of Physics, Columbia University, New York, NY 10027, USA
}%
\author{G.M.~Luke}
\affiliation{%
Department of Physics and Astronomy, McMaster University, Hamilton, Ontario, 
L8S 4M1, CANADA
}%
\date{}
\begin{abstract}
We investigate the N\'eel temperature of Sr$_2$CuO$_3$ as a function of 
the site dilution at the Cu ($S=1/2$) sites with Pd ($S=0$), utilizing 
the muon spin relaxation ($\mu$SR) technique. The N\'eel temperature, which is 
$T_N=5.4$K for the undoped system, becomes significantly reduced for less
than one percent of doping Pd, giving a support for the previous proposal 
for the good one-dimensionality. 
The Pd concentration dependence of the N\'eel temperature is compared with 
a recent theoretical study (S. Eggert, I. Affleck and M.D.P. Horton, 
{\em Phys. Rev. Lett.} {\bf 89}, 47202 (2002)) of weakly coupled 
one-dimensional antiferromagnetic chains of $S=1/2$ spins, and a
quantitative agreement is found. The inhomogeneity of the ordered moment 
sizes is characterized by the $\mu$SR time spectra. We propose a model 
that the ordered moment size recovers away from the dopant $S=0$ sites with
a recovery length of $\xi\approx 150-200$ sites.
The origin of the finite recovery length $\xi$ for the gapless $S=1/2$ 
antiferromagnetic chain is compared to the estimate based on the effective
staggered magnetic field from the neighboring chains. 
\end{abstract}
\maketitle
\section{Introduction}
The discovery of high-$T_c$ cuprates has prompted theoretical and experimental
investigations of low-dimensional spin systems with spin quantum number 
$S=1/2$. There were a series of neutron diffraction studies reported in the 
high-$T_c$ cuprate La$_{2-x}$Sr$_x$CuO$_4$
\cite{KeimerPRL91,KeimerPRB92}, in which the dynamical spin correlation 
was measured. It was found that the inverse correlation length 
$\xi^{-1}(x,T)$ may be divided into the two terms: the temperature 
independent term $\xi^{-1}(x,0)$ 
which is purely determined by the doping concentration and 
the doping independent term $\xi^{-1}(0,T)$ which follows the universal 
temperature dependence. In samples with N\'eel ordering, the 
correlation length diverges $\xi^{-1}\rightarrow 0$ at $T_N$, 
exhibiting long-range magnetic order. 
Recently, effect of static site dilution was also
investigated in La$_2$(Cu$_{1-x}$(Mg,Zn)$_x$)O$_4$ \cite{VajkSCIENCE02}. 
The N\'eel temperature was found to disappear at the 
classical percolation threshold $p_c=0.407$, but the site-dilution dependence
of $T_N$ does not follow the mean-field calculation. The reduction of $T_N$,
 spin-stiffness $\rho_s$ and the equal-time correlation length $\xi(x,T)$ 
were compared with the microscopic quantum mechanical calculations of the 
$S=1/2$ Heisenberg model on a square lattice \cite{VajkSCIENCE02}.

The localized holes in the lightly doped La$_{2-x}$Sr$_x$CuO$_4$ 
and the impurities in the site dilution in La$_2$(Cu$_{1-x}$(Mg,Zn)$_x$)O$_4$ 
break the translational symmetry.
The ordered moment size of these systems should not be homogeneous in space.
However, the signature of the spatial inhomogeneity of 
the ordered moment sizes in cuprates is difficult to detect in the 
momentum space, 
because the doped sites are randomly distributed, the spin quantum 
number $S$ is small and the dimensionality is low. Consequently, the
length scale relevant to the moment size distribution in the impurity 
doped cuprates has not been resolved by the previous neutron 
diffraction measurements. 

Site dilution in the quasi one-dimensional spin systems exhibits more 
important features of the quantum spin systems. 
One peculiar effect of site dilutions in one-dimensional spin systems is 
the creation of a N\'eel order out of the singlet ground state, 
as discovered in the 
impurity doped spin-Peierls material CuGeO$_3$ \cite{HasePRL93b,MasudaPRL98} 
and the 2-leg spin ladder material SrCu$_2$O$_3$ \cite{AzumaPRB97}.
These discoveries have promoted the idea 
that the N\'eel state appears as a competing phase to the original
spin-gapped singlet state of either spin-Peierls or the spin-ladder.
As a result of competition, there exists a length scale which determines 
the spatial variation of the 
ordered moment sizes \cite{FukuyamaJPSJ96a,FukuyamaJPSJ96b}. 
In doped spin Peierls compounds,
the spin decay length in the N\'eel ordered state is determined by the 
ratio between the spin-gap magnitude $\Delta$ and the intra-chain 
antiferromagnetic coupling $J$, such as $\xi/a=J/\Delta$ 
\cite{FukuyamaJPSJ96a}. 
This suggests that the spatial inhomogeneity of the ordered moment size is 
a feature of the spin-gap, and may not appear in the regular quasi 
one-dimensional antiferromagnets with N\'eel order, which are gapless 
($\Delta=0$) and the correlation length diverges.

Recently, a series of measurements of Cu benzoate, a compound with 
$S=1/2$ one-dimensional chains, have identified the existence of an energy 
gap induced by the external magnetic field \cite{DenderPRL97,AsanoPRL00}. 
The origin of the 
energy gap has been interpreted as the effect of the staggered 
magnetic field induced on the Cu sites by the combination of the anisotropic 
$g$-tensor and the external field \cite{AffleckPRB99}. 
This observation points out that the $S=1/2$ spin chain, which has a singlet
ground state and gapless excitations if isolated, may acquire a spin-gap 
under a certain type of perturbation. In quasi one-dimensional 
antiferromagnets, the most common perturbation to an isolated chain is the 
existence of inter-chain interactions $J'$. In the mean-field approximation,
the inter-chain interaction induces an effective staggered magnetic field 
$B\approx zJ'\langle s_z\rangle$, where $\langle s_z\rangle$ and $z$ are the 
magnetic order parameter and the number of the nearest neighboring chains, 
respectively. The inter-chain interaction causes
the N\'eel order \cite{SchulzPRL96} which is gapless because of the 
translational symmetry of the system. With the impurity doping to the $S=1/2$ 
spin-chain, the translational symmetry is broken and the hidden features
of the effective staggered fields from the inter-chain interaction may 
appear in the form of the recovery length $\xi$ of the ordered moment size.
However, this problem is still an open question.

The site dilution should have a destructive effect in the long-range 
coherence of the N\'eel order, especially in low dimensions.
The true one-dimensional chain becomes fragments of finite sizes upon 
impurity doping, and the long-range order becomes impossible. 
However, with the existence of inter-chain interactions, the destruction
of magnetic ordering with site dilution may be moderated; the true 
disappearance of the N\'eel order may occur only at the percolation 
threshold for the three-dimensional lattice structure 
of the inter-chain interactions. Recently, Eggert {\it et al.} presented 
a theoretical estimate for the N\'eel temperature \cite{EggertPRL02} as a 
function of the site dilution 
in the $S=1/2$ quasi one-dimensional antiferromagnet. The predicted $T_N$ 
exhibits a simple reduction as a function of the average chain length $L$. 
Eventhough the ground state of an isolated $S=1/2$ chain is a 
singlet, there was no enhancement of the N\'eel temperature 
upon impurity doping, in contrast with the gapped $S=1/2$ systems, such as 
the spin-Peierls systems and the spin-ladders. 

In order to experimentally investigate the theoretical predictions about 
$T_N$, and the uniformity of the ordered moment sizes in the depleted 
spin-chains, we have performed an investigation of Pd-doped Sr$_2$CuO$_3$. 
The cuprate Sr$_2$CuO$_3$ has received an attention as a model material 
of the $S=1/2$ quasi one-dimensional antiferromagnet; its low N\'eel 
temperature $T_N=5.4$K \cite{KojimaPRL97a} and the large in-chain interaction 
$J\sim 2200$K \cite{AmiPRB95,EggertPRB96,MotoyamaPRL96} suggests its good 
one-dimensionality. The ordered moment size $\approx 0.06 \mu_B$ has been 
obtained by neutron scattering and $\mu$SR measurements \cite{KojimaPRL97a}.
This value is strongly reduced from the full moment size ($=1 \mu_B$) expected 
for the $S=1/2$ spins. The suppressed moment size of Sr$_2$CuO$_3$ follows the 
prediction based on the ``chain mean-field'' theory \cite{SchulzPRL96}, which 
employs the rigorous results of the isolated chains and includes the weak 
inter-chain interaction as the mean-field. 
As the non-magnetic impurity at the Cu site, Zn or Mg substitutions are the 
first choice, as has been already performed in the two-dimensional cuprates 
\cite{VajkSCIENCE02}. However in Sr$_2$CuO$_3$, Zn or Mg ions does not go into
the Cu site. We have employed Pd ion instead; Sr$_2$CuO$_3$ has an 
isostructural compound Sr$_2$PdO$_3$, in which 
the Pd$^{2+}$ ions are in the low-spin ($S=0$) state 
\cite{Sr2PdO3}. This compound has enables us to investigate the non-magnetic 
impurity doping to the $S=1/2$ antiferromagnetic spin-chain.

The structure of this paper is as follows. In section \ref{sec:result}, we
present the magnetic susceptibility data and the result of muon spin 
relaxation ($\mu$SR) measurement of Sr$_2$(Cu$_{1-x}$Pd$_x$)O$_3$. 
The N\'eel temperature was estimated from
the temperature dependence of the muon relaxation rate. In section 
\ref{sec:discussion}, we calculate the magnetic field distribution expected for
the depleted spin-chains with N\'eel order. We assume a zero-moment at the 
impurity site and a recovery length $\xi$ to describe the recovery of the 
ordered moment size into the bulk chain. 
The calculated field distribution is Fourier transformed to obtain the
$\mu$SR spectrum, and employed in the analysis of the nominally pure as well 
as the doped samples. Conclusions are presented in Section 
\ref{sec:conclusion}. 

\section{Experimental results}
\label{sec:result}
We grew single crystals of Sr$_2$(Cu$_{1-x}$Pd$_x$)O$_3$, employing the
traveling-solvent floating-zone technique, with CuO as the solvent. 
Stoichiometric ratio of SrCO$_3$, CuO and PdO powders are prepared, 
mixed in a mortar for an hour and pre-fired in air 
at 900$^\circ$C in an Al$_2$O$_3$ crucible. The powder sample is again mixed 
and fired in air at 950$^\circ$C, before formed into a pressed rod using a 
rubber tube and a water static press. The polycrystalline rod is fired on a 
Pt plate at 1050$^\circ$C. It is important to harden the rod by firing at the 
highest temperature possible for a stable growth of single crystal in the 
floating-zone furnace. We employed a gold-mirror bi-focus furnace made by NEC. 

\begin{figure}
\includegraphics[width=\columnwidth]{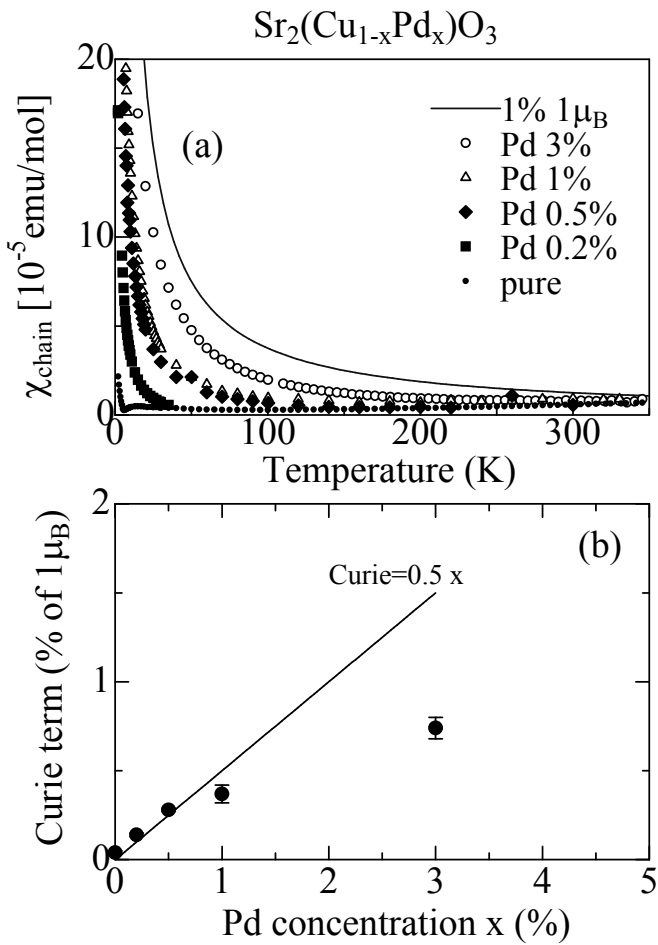}
\caption{\label{fig:chi} (a) Magnetic susceptibility of single crystalline
Sr$_2$(Cu$_{1-x}$Pd$_x$)O$_3$, with the magnetic field applied parallel to the
longest crystallographic axis. (b) Pd concentration dependence of
the Curie term. We note that the substitution with one Pd ion creates
half an impurity moment assumming tht the induced moment size is 1$\mu_B$.}
\end{figure}

The magnetic susceptibility of Sr$_2$(Cu$_{1-x}$Pd$_x$)O$_3$ crystal is shown 
in Fig.~\ref{fig:chi}a, with the magnetic field applied parallel to the 
longest crystallographic axis. An increase of the Curie-Weiss component was
observed as the Pd concentration increases. We assume the conventional form 
for the magnetic susceptibility:
\begin{eqnarray}
\label{eq:chi}
\chi(T) &=& \frac{C}{T+\Theta_W}+\chi_0
\end{eqnarray}
where $C$ is the Curie term, $\Theta_W$ is the Weiss temperature and $\chi_0$
describes the temperature independent susceptibility as a sum of Van Vleck 
paramagnetism and core diamagnetism. 

\begin{table}
\caption{\label{table:chiparam} Parameters of susceptibility and $\mu$SR}
\begin{tabular}{c|c c c c}
\hline\hline
Pd $x$ (\%) & Curie (\%) & $\Theta_W$ (K) & $\chi_0$ (emu/mol) &$\lambda/\xi$\\
\hline
0   & 0.04(1) & 1.7 & -1.8$\times 10^{-5}$  &7.2\footnote{obtained from an analysis of $\mu$SR spectrum (see Discussion).}\\
0.2 & 0.14(1) & 0.8 & -1.1$\times 10^{-5}$  &(2.1)\footnote{numbers in parenthesis are estimates from the Curie term.}\\
0.5 & 0.28(1) & 0.03 & -1.2$\times 10^{-5}$ &(1.0)\\
1.0 & 0.37(5) & 0.33 & -1.3$\times 10^{-5}$ &(0.78)\\
3.0 & 0.74(6) & 2.1 & -1.2$\times 10^{-5}$  &(0.39)\\
\hline
\end{tabular}
\end{table}

The parameters to describe susceptibility are shown in Table 
\ref{table:chiparam}. The Curie term is shown as the concentration of the
impurity moments which are assumed to be 1$\mu_B$.
The calculated concentrations of impurity moments are about {\it half} 
of the doped Pd concentration, as shown in Fig.~\ref{fig:chi}b. 
This Pd concentration dependence of the Curie term may be explained as follows:
One Pd ion creates one chain fragment. Assuming that the intra-chain 
interaction $J$ is much larger than the interaction between the chain 
fragments, the total spin quantum number of the chain fragment is well 
defined: it takes either $S=0$ or $S=1/2$ value, depending on the length of 
the fragment being an even or odd number of spin sites, respectively 
\cite{EggertPRL02}. 
The observed Pd concentration dependence of the Curie term is consistent
with the idea that it originates from the total spin of the created chain 
fragments.
The Weiss temperature $\Theta_W$ exhibits a non-monotonic dependence on the 
Pd concentration. The origin of this dependence is unknown, however, its 
temperature scale is at most $\Theta_W\sim 2$~K, which is being significantly 
smaller than the intra-chain interaction $J\sim 2200$K. This suggests that 
the interaction between the chain fragments are negligible compared to the 
intra-chain interaction, satisfying the condition assumed in the theory 
\cite{EggertPRL02}, which is compared with our measurement in the discussion 
section.

\begin{figure}
\includegraphics[width=\columnwidth]{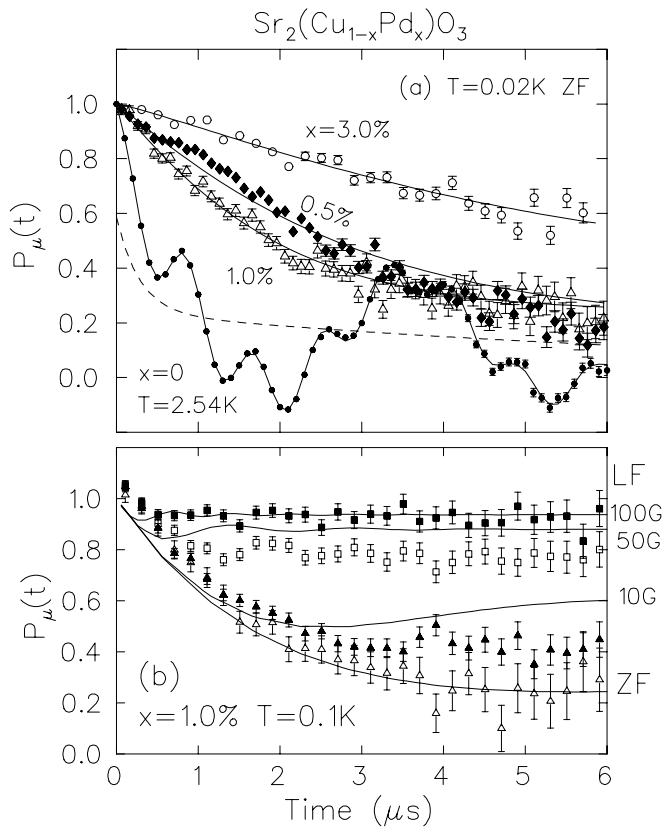}
\caption{\label{fig:muSR} (a) Zero-field muon spin relaxation in 
Sr$_2$(Cu$_{1-x}$Pd$_x$)O$_3$. The solid lines for the doped samples are the
fits using eq. (\ref{eq:exponential}). On the spectrum of the samples with 
$x=0$, the solid line is a plot of eq.(1) in Ref.\cite{KojimaPRL97a}. 
The dashed line is the sum of two exponential functions to describe the 
background relaxation.
(b) Muon spin relaxation in longitudinal field ($H\parallel P_\mu(t)$) for
the $x=1.0$\% sample. The solid lines are the Lorentzian functions in the 
longitudinal fields used in the analysis.}
\end{figure}

It is known that the N\'eel order of Sr$_2$CuO$_3$ is not detectable by 
magnetic susceptibility \cite{MotoyamaPRL96}. This is probably because of 
the very small ordered moment size $\approx 0.06\mu_B$. This feature of the 
material requires employing the muon spin relaxation ($\mu$SR) technique,
which has the highest sensitivity among other experimental techniques to 
detect the magnetic order with small and/or dilute magnetic moments. 
We performed zero-field muon spin relaxation measurement on 
Sr$_2$(Cu$_{1-x}$Pd$_x$)O$_3$ crystals at M15 beam line of TRIUMF (Vancouver, 
Canada).
Muons with 100\% spin polarization were injected into the single crystalline 
sample 
with the initial polarization parallel to the longest crystallographic axis.
This geometry is the same as the one employed in the previous measurement
of nominally pure Sr$_2$CuO$_3$ \cite{KojimaPRL97a}. The time evolution of
muon spin polarization in zero-field is shown in Fig.~\ref{fig:muSR}a. 

In undoped Sr$_2$CuO$_3$, muon spin precession was observed as a consequence 
of  N\'eel ordering \cite{KojimaPRL97a}. 
This indicates that the ordered moment size is relatively homogeneous
in space, and the local field at the muon sites is well defined. 
Upon Pd doping, the muon spin precession disappears and the time evolution 
of muon spin is dominated by the relaxation which has an approximately 
exponential behavior as a function of the time. 
Such exponential relaxation signal in zero-field comes from the $1/T_1$
relaxation caused by spin fluctuations, or alternatively from static fields 
of spatially distributed magnetic moments. The dynamic and static
situations for the muon spin relaxation may be distinguished by the 
``decoupling'' measurements under longitudinal fields applied
parallel to the initial muon spin polarization\cite{HayanoPRB79}. 
The results are shown in Fig.~\ref{fig:muSR}b for the $x=1.0$\% specimen,
together with the analysis using the static relaxation in Lorentzian field 
distribution \cite{UemuraPRB85}. It is clear that the 
relaxation is caused by the static field distribution, as is evident from the 
time-independent behavior in the long terms \cite{UemuraPRB85}. 
This leads to the conclusion that the muon spin relaxation is caused by 
a magnetic order. 

\begin{figure}
\includegraphics[width=\columnwidth]{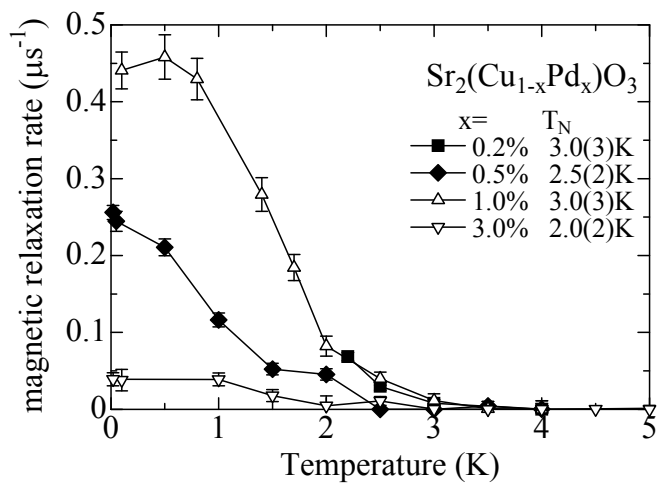}
\caption{\label{fig:rlx} Temperature dependence of the exponential relaxation
rate for Sr$_2$Cu$_{1-x}$Pd$_x$O$_3$. The N\'eel temperatures are estimated
from the temperature points where the relaxation rate start to rise.}
\end{figure}

We analyze the zero-field muon spin relaxation $P_\mu(t)$ in the doped samples
utilizing the phenomenological function:
\begin{eqnarray}
\label{eq:exponential}
P_\mu(t) &=& \exp(-\frac{1}{2}(\Delta_{nd} t)^2)\exp(-\lambda_{mag} t)
\end{eqnarray}
as shown by the solid lines in Fig.~\ref{fig:muSR}a. In the analysis, 
the Gaussian relaxation was assumed to describe the static nuclear 
dipolar-fields, and was set to $\Delta_{nd}=0.12\mu$s$^{-1}$ independent
to the temperature and the sample.
The exponential relaxation rate $\lambda_{mag}$, which parametrizes the muon 
spin relaxation caused by the (atomic) magnetism, are plotted 
in Fig.~\ref{fig:rlx} as a function of temperature. The N\'eel temperature
was defined as the temperature at which the exponential relaxation rate 
start to increase; the estimated N\'eel temperatures are shown in the figure. 

\section{Discussion}
\label{sec:discussion}
\begin{figure}
\includegraphics[width=\columnwidth]{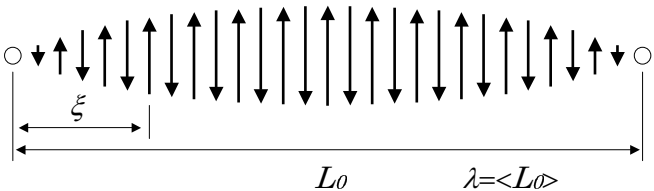}
\caption{\label{fig:spin} Spatial evolution of the magnetic moment size
assumed on a chain fragments. The small circles are the doped Pd sites, 
which were assumed to behave as an $S=0$ impurity. }
\end{figure}

The absence of muon spin precession in Pd doped samples indicates the 
fragility of the spatially homogeneous ordered moment sizes in Sr$_2$CuO$_3$.
With only less than one percent 
doping of Pd, the coherent precession of muon spin disappears, indicating 
that the ordered moment sizes have a broad spatial distribution under site
dilution. The distribution of the moment sizes may be broader than in Zn- or 
Mg-doped CuGeO$_3$, where muon spin precession was observed in the N\'eel
state together with the exponential relaxation \cite{KojimaPRL97b}. 
In the doped spin-Peierls compound, the distribution of the local field was 
consistent with the model that maximum moment size is induced near the doped
center, and the moment size reduces exponentially with the
coherence length $\approx 10$ spin sites \cite{KojimaPRL97b}. 
In Sr$_2$(Cu,Pd)O$_3$ where the Cu moments are depleted by Pd impurities, 
the maximum moment should be located in the middle of the chain fragments 
as shown in Fig.~\ref{fig:spin}. 
Because of the long coherence length expected for the gapless $S$=1/2 spin 
chain, the effect of the non-magnetic impurity may be extended in a large 
area; the maximum moment size of one chain-fragment might strongly depend 
on the chain length, and may not have a well-defined value in the doped 
system. 

In Ref.\cite{KojimaPRL97a}, the $\mu$SR signal of the nominally undoped 
Sr$_2$CuO$_3$ was analyzed by assuming two muon sites each of which consists 
of one precession signal
and one exponentially decaying signal. The latter was interpreted as $1/T_1$
relaxation of the local field component parallel to the initial muon spin
orientation. The exponentially decaying terms describe the background 
relaxation which exhibits a fast front-end before $\approx 1\mu$s as shown 
by the dashed line in Fig.~\ref{fig:muSR}a. The corresponding 
relaxation rate for the front-end ($\approx 5\mu$s$^{-1}$), however, is too 
large for the residual dynamics in the N\'eel ordered state. 
In this section, we calculate
muon spin relaxation for the inhomogeneous N\'eel order with the existence 
of spin vacancies, from which the ordered moment size recovers exponentially 
with the recovery length $\xi$ as shown in Fig.~\ref{fig:spin}. 
The vacancies are the doped Pd ions or the impurities remaining in the 
nominally pure sample which manifest themselves as the Curie term in the 
magnetic susceptibility (Fig.~\ref{fig:chi}). For the nominally pure 
Sr$_2$CuO$_3$, the Curie term corresponds to the impurity level 
$\approx 0.04$\% which corresponds to 
$L \approx 1/(2\times 0.04\times 10^{-2})\approx 1000$ sites of unperturbed 
spin-chain. 

With the model shown in Fig.~\ref{fig:spin}, the moment size $S(z)$ 
behaves as a function of the distance $z$ from the end for a chain 
fragment with length $L_0$:
\begin{eqnarray}
\label{eq:Sz}
S(z) &=& s_0\left\{ 1-\frac{\cosh((z-\frac{1}{2}L_0)/\xi)}{\cosh(\frac{1}{2}L_0/\xi)}\right\}
\end{eqnarray}
where $s_0$ is the moment size at infinity ($s_0\approx 0.06\mu_B$)
and $\xi$ is the moment-size recovery length. By assuming that the muon 
local field $H$ is proportional to the nearest moment size 
($H(z)\propto S(z)$: local moment density approximation), the distribution 
function $\rho(H;L_0)$ of the local fields be obtained by the density of 
the states for the field $H$ along the chain fragment $L_0$: 
\begin{eqnarray}
\label{eq:rhoH}
\rho(H;L_0)&=&\frac{dz}{dH(z)}.
\end{eqnarray}
In the doped material, the impurity site should be located randomly. 
The experimentally observed field distribution function $\rho(H)$ in this
situation is the average of $\rho(H;L_0)$ for the Poisson distribution of 
the chain length $L_0$. The same procedure was taken for the model relaxation
employed in the analysis of the $\mu$SR spectra of doped CuGeO$_3$ \cite{KojimaPRL97b}. 

\begin{figure}
\includegraphics[width=\columnwidth]{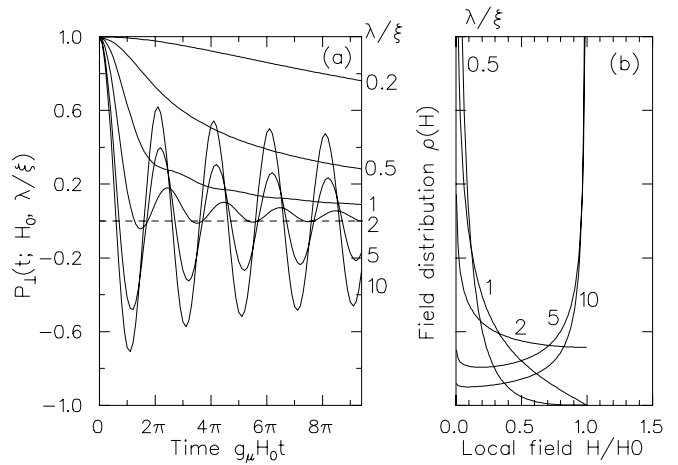}
\caption{\label{fig:theory} (a) The muon relaxation function and (b)
local-field distribution for the inhomogeneously N\'eel ordered spin-chain 
(Fig.~\ref{fig:spin}).}
\end{figure}

The model muon relaxation function and the corresponding local field 
distribution function for the inhomogeneous ordered moment size 
(Fig.~\ref{fig:spin}) are shown in Fig.~\ref{fig:theory}. The behavior of the
distribution function $\rho(H)$ is characterized
by the ratio between the average chain-length 
$\lambda=\langle L_0\rangle\approx 1/x$ and the 
moment-size recovery length $\xi$. In comparison to the model for 
doped CuGeO$_3$ (Fig.~4b of Ref. \cite{KojimaPRL97b}), the local field
distribution (Fig.~\ref{fig:theory}b) has the same shape
but the field axis is reversed between $H=0$ and $H_0$. 
This is because in the doped CuGeO$_3$, impurities create magnetic 
moments, whereas in Sr$_2$CuO$_3$ moments are depleted from the chain. 
The role of the zero-field $H=0$ and the maximum field $H=H_0$ 
are reversed between CuGeO$_3$ and Sr$_2$CuO$_3$ cases. 
As shown in Fig.~\ref{fig:theory}a, when the ratio 
$\lambda/\xi\gg 1$, there is a coherent precession of muon spins because 
the moment size is relatively homogeneous and the local field distribution 
exhibits an isolated peak at $H=H_0$. As the ratio approaches unity, 
the precession amplitude is diminished and at $\lambda/\xi=1$, 
only a weak bump remains at the position of the first precession peak.
Below $\lambda/\xi=0.5$, there is no visible oscillation and 
the relaxation is mostly exponential with time. 
Compared to the model for doped CuGeO$_3$ (Fig.~4b of Ref.\cite{KojimaPRL97b})
with the same level of background relaxation, the amplitude of the precession 
signal component is much smaller in this model for depleted chains.  
The reason for the difference stems from the maximum moment size
of the chain fragments: in the model of doped CuGeO$_3$, it was assumed
that the ends of all chains have the same induced moment size $s_0$. 
In this model for depleted chains, the maximum moment size appears at the 
center of the chain fragments, and depends on their length. 

\begin{figure}
\includegraphics[width=\columnwidth]{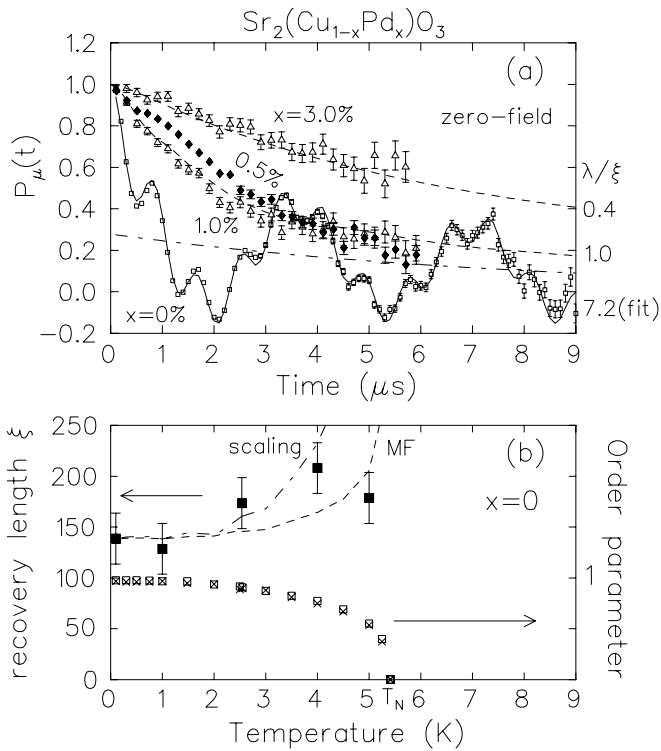}
\caption{\label{fig:theoryfit} (a) Muon spin relaxation in Sr$_2$CuO$_3$ 
analyzed with the relaxation function eq.(\ref{eq:Pmu2}) (solid line).
The dot-dashed line is the contribution from the parallel-field term 
($A_\parallel^{A+B}$) introduced in the analysis. Dashed lines are 
eq.(\ref{eq:Pmu2}) with the same parameters as in the $x=0$ case, except the 
$\lambda/\xi$ ratio. 
(b) Temperature dependence of the recovery length
$\xi$ and the magnetic order parameter as derived from the analysis. 
The dashed and the dot-dashed lines are the theoretical temperature 
dependence of $\xi$ calculated from the experimental order parameter
(see section \ref{sec:discussion}).}
\end{figure}

If we represent the relaxation function shown in Fig.~\ref{fig:theory}a 
with a symbol $P_\perp(t;H_0,\lambda/\xi)$, the muon relaxation function
for one muon site becomes:
\begin{eqnarray}
\label{eq:Pmu1}
P_\mu(t) &=& A_\perp P_\perp(t;H_0,\lambda/\xi)+ A_\parallel\exp(t/T_1),
\end{eqnarray}
where $A_\perp$ ($A_\parallel$) is the amplitude of the local field 
component which is perpendicular (parallel) to the initial muon spin 
polarization, and $T_1$ is the spin-lattice relaxation
time for the parallel component. 
As can be seen from the existence of two precession frequencies
(Fig.~\ref{fig:muSR}a), there are two muon sites in Sr$_2$CuO$_3$ which, most
likely, correspond to the muons attached to the in-chain and out-of-chain
oxygen sites (see Appendix \ref{sec:muonsite}).
We introduced the local fields $H_0^{A}$ and $H_0^{B}$ for the each site, 
and obtained the following phenomenological function for the muon spin 
relaxation:
\begin{eqnarray}
\label{eq:Pmu2}
P_\mu(t) &=& A_\perp^A P_\perp(t;H_0^A,\lambda/\xi)
	+A_\perp^B P_\perp(t;H_0^B,\lambda/\xi)\nonumber\\
         &+& A_\parallel^{A+B}\exp(t/T_1),
\end{eqnarray}
where $A_\perp^A$, $A_\perp^B$ and $A_\parallel^{A+B}$ are the amplitudes 
of each components. Here, the two $1/T_1$ signal components are combined 
in one relaxation amplitude ($A_\parallel^{A+B}$), since the background 
relaxation rate is too small to distinguish the contributions from the two 
sites. For the nominally pure sample, the precession signals
can be analyzed with eq.(\ref{eq:Pmu2}). 
The solid line in Fig.~\ref{fig:theoryfit}a is the result of the fit and the
dot-dashed line is the contribution from the $1/T_1$ relaxation term. 
It is noted that there are no extrinsic parameters introduced to describe 
the damping of the oscillation amplitude nor the early front-end relaxation. 
However, eq.(\ref{eq:Pmu2}) based on the inhomogeneous moment size distribution
(Fig.~\ref{fig:spin}) describes the over-all feature of the muon spin
relaxation fairly well. The ratio between the 
average chain-length $\lambda$ and the recovery length $\xi$ yields
$\lambda/\xi\approx 7.2$ from the analysis. The fast front-end relaxation 
at $t<1\mu$s is also described by the perpendicular terms 
($A_\perp^A$ and $A_\perp^B$). In this analysis, the $1/T_1$ relaxation rate 
is not as large as the one obtained in the previous analysis of 
Ref.\cite{KojimaPRL97a}.

The magnitude of the Curie term in the susceptibility for the nominally
pure Sr$_2$CuO$_3$ (Table \ref{table:chiparam}) suggests the average 
chain-length for this nominally pure sample is 
$\lambda\approx 1/(2\times0.04\%)\approx 1000$ lattices.
This and the ratio $\lambda/\xi=7.2$ obtained from the fitting analysis 
suggests that the recovery length of the moment size is $\xi\approx 150$ 
lattices, which is at least one order of magnitude longer than in the N\'eel 
state of doped CuGeO$_3$ in which $\xi\approx 10$ \cite{KojimaPRL97b}. 
The absence of a nominal spin-gap in the $S=1/2$ antiferromagnetic chain is 
most likely the cause for the correlation length being longer than in the doped
spin-Peierls compound CuGeO$_3$. 
The temperature dependence of the recovery length ($\xi$) and the magnetic
order parameter ($\langle s_z\rangle\propto$ precession frequency) are plotted 
in Fig.~\ref{fig:theoryfit}b. The temperature dependence of $\xi$ 
is weak and possiblly exhibiting a slight increase at higher temperatures.

Based on the magnitude of the Curie terms, one can estimate
the ratio $\lambda/\xi$ for the Pd-doped compounds, which are summarized in
Table \ref{table:chiparam} in parenthesis. 
The model relaxation functions eq.(\ref{eq:Pmu2}) for
$\lambda/\xi= 1$ and 0.4 are shown as 
the dashed lines in Fig.~\ref{fig:theoryfit}a. These two parameter values 
corresponds to the Pd 0.5\% and 3\% doped samples, respectively. 
Since the $\mu$SR spectra of the doped systems do not exhibit the spectral 
features which were present in the $x=0$ sample, it is not possible to 
experimentally determine the parameters in eq.(\ref{eq:Pmu2}). Here we 
assumed the same amplitudes ($A_\perp^A$, $A_\perp^B$ and $A_\parallel^{A+B}$),
local fields ($H_0^A$ and $H_0^B$) and $1/T_1$ relaxation rate
as determined in the $x=0$ sample, and varied the $\lambda/\xi$ parameter. 
This may be a good approximation in low doping in which the orientation of 
the local fields are approximately the same as in the nominally pure case. 
The calculated relaxation functions eq.(\ref{eq:Pmu2}) 
agree well with the behavior of the experimentally obtained $\mu$SR spectra of 
Sr$_2$(Cu$_{1-x}$Pd$_x$)O$_3$ as shown in Fig.~\ref{fig:theoryfit}a. 
Eventhough the $\mu$SR spectra in the Pd-doped samples do not show precession, 
this might originate in the static N\'eel order with spatial distribution of 
moment-sizes. The decoupling measurement (Fig.~\ref{fig:muSR}b) confirms that
the relaxation is static, which is a support to the idea that it originates 
from N\'eel order. Because of the long recovery length $\xi$, the small amount 
of impurity ions disturbs the ordered-moment size in a large area, 
so that the spin precession of the local magnetic probe becomes invisible.

In the $\mu$SR time spectrum of the $x=0.5$\% sample, there is a hump in the 
measured polarization at $\approx 1 \mu$sec (Fig.~\ref{fig:theoryfit}a), 
at the position of the first precession peak for the nominally pure compound.
From this result, it is suggested that
the maximum moment size $s_0$ does not change in the sample with 0.5\% Pd 
doping. For the $x=1.0$\% and 3.0\% samples, such hump disappears, indicating 
that as the doping level increases, the parameter enters to the 
$\lambda/\xi\lesssim 1$ regime. The estimate of the $\lambda/\xi$ value 
(Table \ref{table:chiparam}) confirms this conclusion.

\begin{figure}
\includegraphics[width=\columnwidth]{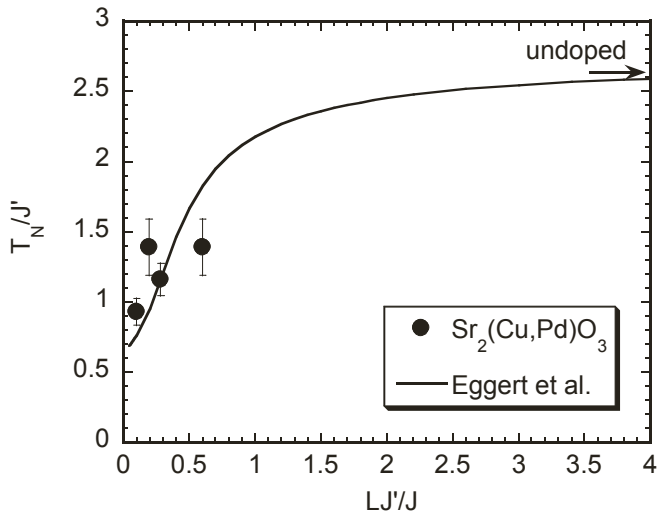}
\caption{\label{fig:TnvsL} N\'eel temperature vs. average chain length. 
The theoretical line is taken from Eggert {\it et al.} \cite{EggertPRL02}.}
\end{figure}

In Fig.~\ref{fig:TnvsL}, the N\'eel temperature of Sr$_2$(Cu,Pd)O$_3$
and the theoretical curve \cite{EggertPRL02} are plotted. We estimate the 
average chain length $L$ of Sr$_2$(Cu,Pd)O$_3$ from the Curie term
of the susceptibility, assuming that one Pd ion creates either $S=0$ or 
$S=1/2$ with the equal probability. 
The horizontal and vertical axes of Fig.~\ref{fig:TnvsL} are normalized by the 
inter-chain interaction $J'$ which has not been experimentally obtained for 
Sr$_2$CuO$_3$. However the ratio $T_N/J = 4\times 10^{-4}$ has been 
estimated \cite{KojimaPRL97a} and the theory proposes $T_N/J'\approx 2.6$ in 
the long chain-length limit (Fig.~\ref{fig:TnvsL}). From these, one can obtain 
the ratio between the inter- and intra-chain couplings 
$J'/J\approx 1.6\times 10^{-4}$. This value was employed to scale the average
chain length $L$ in the horizontal axis of Fig.~\ref{fig:TnvsL}.
The agreement between theory and measurement regarding the N\'eel 
temperature is reasonablly good, suggesting that the theory correctly 
estimates the energy scale of the magnetic order. 

Within the same theoretical framework, the spatial 
distribution of the ordered moment size has been calculated recently
\cite{EggertJMMM04}. The assumption made is that the $S=1/2$ 
antiferromagnetic chain acquires the staggered moment size $\langle s_z\rangle$
in a consistent way from the effective staggered magnetic field $B$
originating from the neighboring ordered chains. The intra-chain interaction 
$J$ propagates the effect of the chain-end, reducing the 
ordered moment size at its vicinity. The inter-chain interaction $J'$
recovers the ordered moment size by inducing the staggered magnetic field 
from the neighbouring chains. The competition between these two interactions
on the chain fragments causes the spatial
distribution of ordered moment sizes as shown in Fig.~ \ref{fig:spin}. 
A scaling argument proposes that the recovery length $\xi$ is given by
$\xi\propto B^{-2/3}$ \cite{EggertPC04,EggertJMMM04}, where $B$ is the 
effective staggered field. In the mean-field approximation, the staggered
field can be written as $B=zJ'\langle s_z\rangle$, where $J'$ is the 
inter-chain interaction, $z$ is the number of neighboring chains and 
$\langle s_z\rangle$ is the magnetic order parameter. Combining these two
relations, the recovery length in the mean-field approximation is given by
$\xi_{MF}\propto \langle s_z\rangle^{-2/3}$. 
There is also a scaling relation between the magnetic order parameter and the 
staggered field: $\langle s_z\rangle\propto B^{1/3}$ \cite{EggertJMMM04}. 
With this and the scaling relations relation between $\xi$ and $B$, 
one can obtain a scaling prediction for the  
recovery length: $\xi_{SC}\propto \langle s_z\rangle^{-2}$. 
In either case, the temperature dependence of the recovery length is set by 
that of the order parameter $\langle s_z \rangle$ but with a different 
exponent. 

The temperature dependence of the recovery length in the mean-field 
approximation ($\xi_{MF}$) and the scaling result ($\xi_{SC}$) 
are shown in Fig.~\ref{fig:theoryfit}b, based on the experimentally 
obtained magnetic order parameter. Since we do not know the over-all scale 
factor of the recovery length, we employed the experimental value obtained 
at low temperature. The experimental result is more consistent with the full 
scaling result $\xi_{SC}$, rather than with the value expected in the 
mean-field framework, except at the vicinity of the N\'eel temperature. 
The critical exponent of the order parameter as a function of temperature
($\beta\approx 0.2$)\cite{KojimaPRL97b} also indicates that the system is
in the scaling regime rather than in the three-dimensional mean-field
regime in which $\beta\approx 0.5$. 
Our measurements suggest that the finite recovery length of 
the ordered moment sizes is determined by the effective staggered magnetic 
field $B$ originating from the neighboring chains. The over-all scale factor 
$\xi/a\approx 150$ in the low temperature limit is yet to be calculated
theoretically for the parameters of Sr$_2$CuO$_3$. 

In the theoretical calculation of chain fragments with the lengths  
longer than the recovery length ($L_0/\xi\gtrsim 1$), 
the self-consistent moment size exhibits
a two peak structure in the distribution function 
(Fig.~1 of Ref.\cite{EggertJMMM04}). One peak is located almost at zero-moment 
size and the other appears at a slightly larger size than that for the 
unperturbed chain. The former and the latter peaks originate from the chain 
fragments with even and odd number of spins, respectively \cite{EggertJMMM04}.
We have found 
that the Fourier transformation of the moment size distribution function
exhibits clear precession, as a result of the isolated peak for 
the odd-numbered fragments. This theoretical result is in contrast to the 
experimental observation (Fig.~\ref{fig:muSR}a), in which the precession is 
dampled upon doping. There are two possible explanations: (1) The moment size 
of the odd length chains may be reduced due to quantum fluctuations,
as has already been suggested in Ref. \cite{EggertJMMM04}. 
(2) The sum of dipolar fields at the muon site contains contributions from 
neighboring chains, smearing out the even-odd effect. The proposed muon 
sites and the dipolar field calculations (Appendix \ref{sec:muonsite}) 
demonstrate that the contribution from the nearest neighbor chain is 
actually dominant, at least for the higher frequency site. 
This suggests that the quantum fluctuation scenario (1)
may be a more favoured explanation for the absence of muon spin precession in 
Sr$_2$(Cu,Pd)O$_3$.

\section{Conclusions}
\label{sec:conclusion}
We have investigated how N\'eel order is destructed with non-magnetic 
impurity doping in the quasi one-dimensional $S=1/2$ antiferromagnet. The
model material employed is Sr$_2$(Cu$_{1-x}$Pd$_x$)O$_3$. 
The susceptibility at $T>T_N$ exhibits Curie-Weiss behavior. The magnitude 
of the Curie term is consistent with a model in which the creation of 
half an induced moment (size = 1$\mu_B$) occurs due to substitution with one
Pd ion. This suggests that the induced paramagnetic moment originates from 
the total spin of the chain fragments which is either $S=0$ or $S=1/2$
depending on the fragment consists of even or odd number of spin sites. 
The muon spin relaxation of the nominally pure Sr$_2$CuO$_3$ was re-analyzed 
with a model for the spatially inhomogeneous N\'eel ordered state. 
The length scale 
$\xi$ was introduced to describe the recovery of ordered moment size away 
from the spin defect. It was suggested that the length scale $\xi$ originates
from the effective staggered magnetic field of the neighboring chains. 
The length scale $\xi\approx 150$ lattices 
for Sr$_2$CuO$_3$ is more than 10 times longer than in doped CuGeO$_3$, 
which reflects the nominally gapless characteristic of the system. 

In the Pd doped samples, the absence of muon spin precession for 
less than 1\% doping level is consistent with a large recovery length $\xi$.
The N\'eel temperature, which was defined as the temperature where muon
spin relaxation rate starts to increase, exhibits a good agreement with 
recent theoretical calculation. 

\section*{Acknowledgement}
The authors would like to thank Dr. S.~Eggert and Prof. I.~Affleck
for stimulating discussions. We also thank Prof. Y.~Kato for valuable 
comments. 
The research in this paper has been financially supported
by NEDO International Joint Research Grant and by
COE \& Grant-in-aid for Scientific Research from Monbusho.
Work at Columbia University was supported by NSF-DMR-01-02752
and NSF-CHE-01-17752 (Nano-scale Science and Engineering
Initiative).

\appendix
\section{Muon sites}
\label{sec:muonsite}
The crystal structure of Sr$_2$CuO$_3$ has two non-equivalent oxygen sites
\cite{AmiPRB95}. Site O(1) is out of the chain oxygen site of the 
corner-shared CuO$_3$ plaquett, and site O(2) is the shared oxygen of the 
two neighboring plaquetts forming the chain structure. 
In cuprates, it has been
proposed that muons form an O-$\mu^+$ bond with oxygen ions in an analogy with 
the hydrogen bonding \cite{HittiHI90}. In this appendix, we calculate 
the electro-static potential for a muon which was assumed to form an O-$\mu$ 
bond with a bond length of 1\AA\ and determine the orientation of the bond 
with respect to the crystallographic axis. We also calculate the dipolar 
fields and compare with the experimentally observed local fields.

\begin{figure}
\includegraphics[width=\columnwidth]{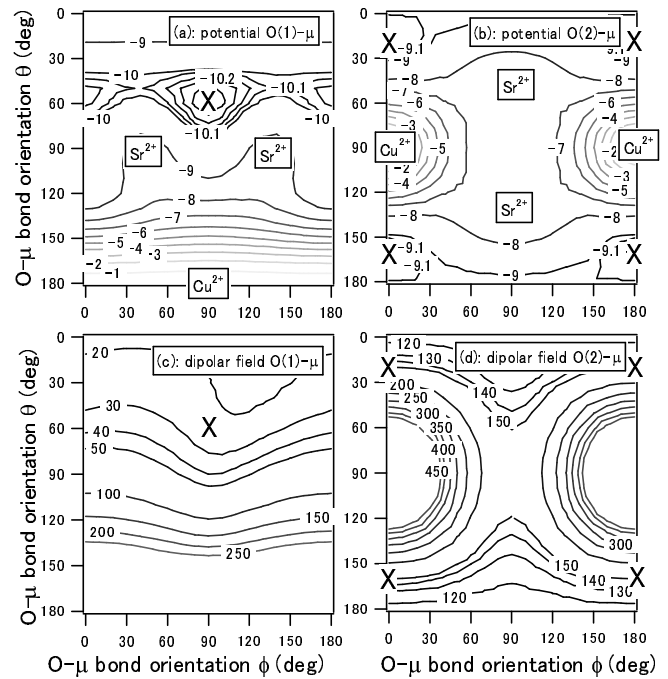}
\caption{\label{fig:esp-dip} (a) and (b): Electro-static potential for a muon
1\AA\ away from the oxygen site O(1) and O(2), respectively. The unit of the
potential is in eV. The crosses are the potential minima which are the most
probable orientation for the O-$_\mu$ bond. (c) and (d): Dipolar fields 
calculated for the same O-$\mu$ bonding. The unit of the field is in Gauss 
for the spin structure determined by the neutron diffraction measurement
\cite{KojimaPRL97a}.}
\end{figure}

\begin{figure}
\includegraphics[width=\columnwidth]{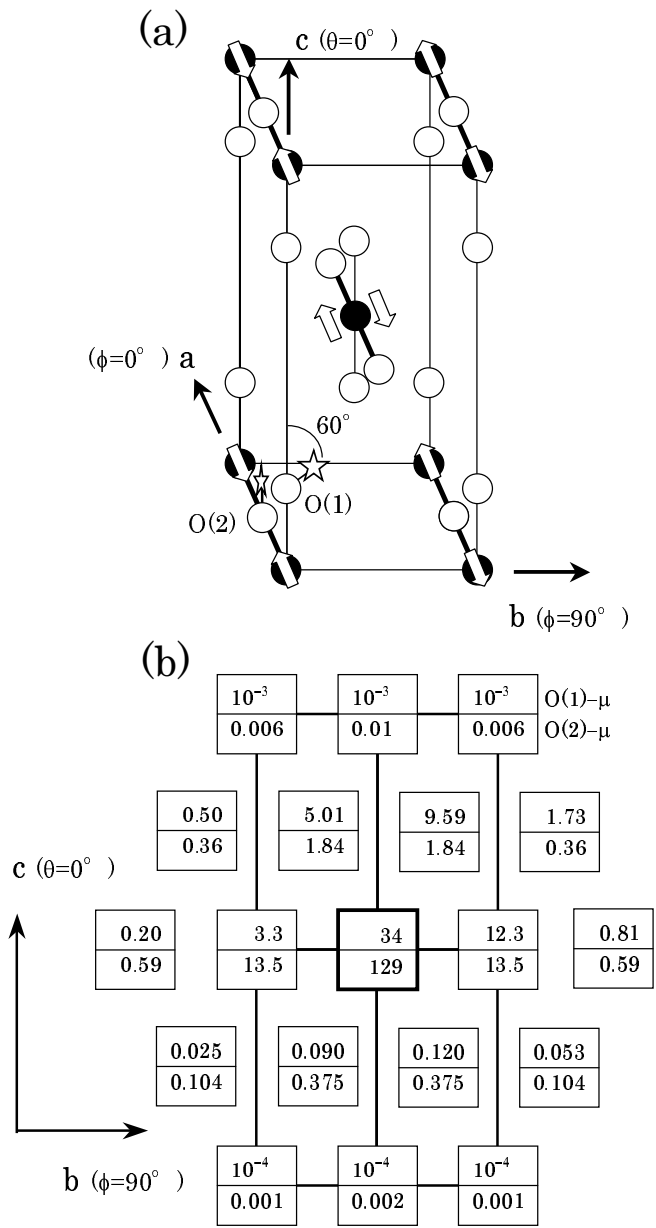}
\caption{\label{fig:site-dip} (a) The crystal structure of Sr$_2$CuO$_3$.
O(1)-$\mu$ and O(2)-$\mu$ sites as determined by the electro-static potential
calculation are shown by the star symbols which are attached to the each 
oxygen sites. (b) The dipolar fields from the each chains to the muon sites.
The boxes are distributed at the position of the chains seen along the
$a$-axis. The upper and lower panels of the boxes are the dipolar
fields for O(1)-$\mu$ and O(2)-$\mu$ sites, respectively. 
The unit of the fields is in Gauss.}
\end{figure}

In Fig.~\ref{fig:esp-dip}a and \ref{fig:esp-dip}b, contour plots of the 
electro-static potential are drawn for the muons forming the O-$\mu$ 
bonds with the O(1) site and O(2) site, respectively. 
The electro-static potential is calculated using the 
method of Ewald sum \cite{KittelIntro} assuming the formal point charges 
located at the ionic positions. The minimum of the potential exists at 
$(\theta,\varphi)_{(1)}=(60^\circ,90^\circ)$ and 
$(\theta,\varphi)_{(2)}=(20^\circ,0^\circ)$ for the 
O(1)-$\mu$ and O(2)-$\mu$ bonds, respectively, which are shown by the cross 
symbol in the figures. The polar coordinate is defined 
as the $\theta=0$ direction being parallel to the longest crystallographic 
axis ($c$-axis), and the $\varphi=0^\circ$ direction in the basal plane being 
parallel to the CuO chains ($a$-axis). 
In Fig.~\ref{fig:site-dip}, the muon sites in the real
space crystal structure are shown by the star-symbols. 
The potential minimum value for the bond length of 1\AA\ 
is deeper for the O(1)-$\mu$ site (-10.4eV), than for the O(2)-$\mu$ site
(-9.2eV). This suggests that O(2)-$\mu$ site may have a shorter bond-length 
to gain the electrostatic potential of the oxygen ion. 


The ordered moment size and its orientatin of Sr$_2$CuO$_3$ has been obtained 
by neutron diffraction measurements as 0.06$\mu_B$ pointing along the CuO 
chain direction, respectively \cite{KojimaPRL97a}. The corresponding
dipolar fields for the spin structure obtained are calculated for 
the two kinds of O-$\mu$ bonds. The results are shown in 
Fig.~\ref{fig:esp-dip}c and \ref{fig:esp-dip}d. At the orientation of the 
minimum potential (cross symbols in the figures), the dipolar fields are 
24G and 150G for the O(1)-$\mu$ and O(2)-$\mu$ site, respectively. 
The experimentally observed local fields in Sr$_2$CuO$_3$ are 
23.3G and 97.7G \cite{KojimaPRL97a}. The former is close to the calculation 
of O(1)-$\mu$ site, which is most likely the muon position responsible to the
23.3G signal. The 97.7G signal does not agree well with the calculation for
the O(2)-$\mu$ site. The disrepancy might originate from the bond length
which could be shorter for O(2)-$\mu$ site. We have calculated the 
electro-static potential and the dipolar fields for a shortened O(2)-$\mu$ 
bond length of 0.9\AA\ and found that the dipolar field at the potential 
minimum is reduced to 120G for the same spin structure. Since the pseudo 
minimum position for the O(1)-$\mu$ site $(60^\circ,0^\circ)_{(1)}$ exhibits 
a much smaller dipolar field (40G), the 97.7G signal most likely
originates from the O(2)-$\mu$ site with a shortened bond length.

For the muon sites proposed above, we calculate the contribution of the 
dipolar fields from neighboring chains. In Fig.~\ref{fig:site-dip}b, we
show the magnitude of the dipolar fields from the chains at the position 
of the boxes which represent the CuO chains projected to the $b\times c$-plane.
The upper and lower panels of the boxes indicates the dipolar fields of the 
O(1)-$\mu$ and O(2)-$\mu$ sites, respectively. For the O(1)-$\mu$ site, 
the contribution from the nearest neighboring chain is as large as 1/3 of the 
main contribution. This contribution from the neighboring chains broadens
the local field distribution from that expected by the spatial distribution
of the ordered moment sizes. However, the precession signal with higher
frequency of the nominally pure Sr$_2$CuO$_3$ originates from the O(2)-$\mu$ 
site, at which the contribution of the dipolar field from the neighboring 
chains is at most 1/10 of the main contribution. This calculation 
suggests that the precession signal with higher frequency would survive, 
if the ordered moment size distribution has a distictive peak at a finite 
frequency as proposed in the theoretical calculation \cite{EggertJMMM04}. 
The absense of the 
precession signal in the Pd-doped sample suggests that the isolated
peak of the ordered moment size, which originates from the chain fragments 
with odd number of spins, is actually diminished in Sr$_2$CuO$_3$ due to the
effects which are not included in the theoretical calculation.

\end{document}